**Academia Eğitim Araştırmaları Dergisi**
www.academiadergi.com

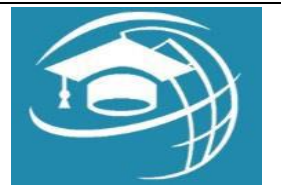

# Kitlesel Açık Online Dersler ve Bulut Bilişim

**Mansur BEŞTAŞ[1]**

1 Siirt Üniversitesi, *Türkiye*

**Özet**

Günümüz bilgisayar teknolojisinin en önemli alanlarından biri Bulut Bilişim'dir. Bulut Bilişimin birçok alanda sağladığı kolaylıklar göz ardı edilemez. Bulut bilişimin sağladığı imkânlardan etkilenen bir alanda eğitim alanıdır. Bulut bilişimin öğrenme alanında sağladığı imkânlar sonucunda MOOC ortaya çıkmıştır. MOOC son yıllarda sürekli olarak artan talep ile e-öğrenme alanında umut vaat eden bir yeri olduğu düşünülmektedir. Bu nedenle MOOC modeli İş modeli yönünden incelenecektir. Müşteri tipleri, vadettiği hizmetler, elde ettiği gelirlerin kaynaklarının neler olduğuna bakılacak karşılaştırma tablo şeklinde verilecektir. Bulut Bilişimin servis modellerinden platform ve yazılım hizmet ile olan ilişkisi ortaya koyulmaya çalışılacaktır. Bu çalışma ile Bulut Bilişim ve MOOC öğrenme metodunun ilişkisini ortaya koymak amaçlanmaktadır.

© 2017 AEAD

**Anahtar Kelimeler**
Açık online dersler, Bulut bilişim, e-öğrenme

# Massive Open Online Courses and Cloud Computing

**Abstract**

One of the most important areas of today's computer technology is Cloud Computing. Benefits of Cloud Computing in various areas cannot be ignored. One field affected by the opportunities provided by Cloud Computing is education. MOOC emerged as a result of opportunities in the field of learning provided by cloud computing. With the ever-increasing demand in recent years, MOOCs are considered to have a promising place in e-learning. Thus, the MOOC model will be investigated as a business model. Customer types, services provided, the sources of income will be analyzed and tabulated. It will be aimed to reveal its relationship with platform and software service among the service models of computing. It was aimed with this study to present the relationship between Cloud Computing and MOOC learning method.

© 2017 AEAD

---

[1] Siirt Üniversitesi, Bilgi İşlem daire Başkanlığı, Siirt-Türkiye, mansur@siirt.edu.tr

# INTRODUCTION

Opportunities are increasing daily in the field of computing. Benefits and opportunities created by these influences present themselves inevitable in every aspect of life. The field of education is one of these. E-learning is defined as the process of achieving the education procedure using technology and computing tools. It may be seen that computing opportunities have been used in various formats through time. Cloud computing satisfies the ever-increasing need for resources today in e-learning. The easier satisfaction of resource needs has made the MOOC model applicable in the field of e-learning. The reason for choosing this subject are is that, as far as we could follow, studies where the MOOC model was discussed were mostly conducted on dangers in general, experience of learning, pedagogics, technology, and perspectives of trends (Liyanagunawardena, Adams, & Williams, 2013). In scope of the study, issues such as which groups constituted the target customers in the MOOC model, what products are supplied to the target audience and sources of income. This research study investigated the most significant suppliers that are using the MOOC model in terms of a business model, and analyzed the relationship based on cloud computing service models. Some of the top websites where the MOOC model is supplied, Coursera, EDX, Udemy and Udacity were included in the study. As a method in the study, an approach was determined based on the research method Giessmann and Slabeva reported in their article (Giessmann & Stanoevska-Slabeva, 2012). As a result of the data obtained by a detailed investigation and data analysis of these sites, it was found that the area in question is still in a process of change and transformation. It was found that the structure is increasingly developing services towards the SaaS service model. Reports on the effects of MOOCs on the educational life were published in the literature (Hollands & Tirthali, 2014). Their effects on higher education and pedagogical aspects were intensively studied. However, it was seen that their cloud computing aspect, which is the source of their fast response to large masses and continuity, was not mentioned or investigated in terms of a business model. It was aimed to fill this dearth found in the literature.

**RESEARCH APPROACH AND METHOD**

This study used the method of business model investigation determined by Giessmann and Slabeva for suppliers active in the cloud computing area (Giessmann & Stanoevska-Slabeva, 2012). For the MOOCs subject to case analyses, which customer layers they have, what they promise to their customers as services, and their sources of income were investigated. This was followed by a summary of the situation with classification tables.

**FINDINGS AND DISCUSSION**

**Cloud Computing**

Cloud computing is defined as the process of running operations in usage of a pooled capacity of computing in any needed area over the internet by using basic skills of information and communication technologies. Its general characteristic is that it quickly and autonomously designs operations management, accessibility and computing resources in a pool based on necessities and supplies these for the usage of all services in a measurable way. Cloud computing is used in various areas due to the opportunities it provides. It may be understood in three categories based on forms of usage (service models). Software as a Service: this means the provision of a piece of software to one or more people or institutions. Gmail and salesforce.com may be given as examples. Platform as a Service: this is a type of service where the service provider provides the opportunity of managing an operating system, a software framework database, or web servers. Examples are Orangespace, Mendix , Heroku. Infrastructure as a Service: this is a type of service where the service provider provides services such as processing power, storage, high speed internet connection and RAM capacity. In this service model, virtual machines are usually provided. Amazon and digitalocean may be given as examples (Mell and Grance, 2011).

**What is E-learning?**

E-learning is supporting the learning process by a digital technology or achieving the entire process using digital technologies (Holmes & Gardner, 2006). This technology is generally a desktop computer, a laptop, a tablet or a smartphone. In the learning process, e-learning is sometimes used as the process itself, while it is used in communicating and analyzing feedbacks in other times (Clark & Mayer, 2016).

**The Size of the E-learning Economy**

The entire e-learning market is predicted to be $ 107 billion by 2015 (Docebo, 2016). According to the global market insight website, in 2015 the entire e-learning market was realized at $ 165 billion (internet, 2016). It is expected that the LMS market in e-learning market will be 4 billion dollars in 2015 (Pappas, 2015). Ambient insight website 2016-2021 says that the self-learning e-learning market has reached $ 46.6 billion for 2016. In 2021 it will be 33.4 billion dollars (Adkins, 2016). Instead of the e-learning market, mobile learning, simulation-based, game-based learning market is predicted to develop.

**What Is the Relationship between Cloud Computing and E-learning?**

We are experiencing a process where education and instruction is even more critical for absolute success in this developing world. As a result of this, all opportunities of technology started to be utilized to improve learning opportunities and satisfy the received demand (Al-Zoube, 2009). While the learning needs of individuals necessary for an average professional skill were very limited in the past, it may be seen that the ready-availability level today is much higher (Clark & Mayer, 2016). Learning continues through life in order to satisfy these needs. Education and computing technologies have important roles in satisfying these needs of people in the area of education. While e-learning tools were limited to blog pages, forums, static web pages and visual learning platforms in the beginning, today, with the development of WEB 2.0 technologies, interactive web pages became widespread and as a result, java script and html 5 technologies developed (Jain & Chawla, 2013). All these led to a more active interaction of the individual and the virtual environment. With the developed technology, it was in evitable that the prepared e-learning objects and learning environment increased in number. This situation led to the development of the e-learning market by increasing towards the needs of individuals. High numbers of users necessitated high amounts of storage space and processing power. With today's technology, this need can only be satisfied cloud computing opportunities. The storage capacity needs of the educational materials provided for services and the need for high bandwidth in usage were solved with the cloud computing service model of IaaS (Infrastructure as a Service). Storage of high numbers of educational materials and user information in databases was solved with the cloud computing model of PaaS (Platform as a Service). Provision of the environment where the software can work for combination of all these with the user was made possible with SaaS (Software as a Service). The examples given regarding the service models of cloud computing are superficial. Reviewing the studies will reveal that cloud computing provides benefits in various areas. It provides massive opportunities for educational audiences with low budgets but good ideas.

**The MOOC (Massive Open Online Course) Model in E-learning**

MOOC is an e-learning model where an unlimited number of students can reach quality learning content from afar and the access can be provided only by an internet connection and a web browser (McAuley, Stewart, Siemens, & Cormier, 2010). Its first examples emerged in 2008. It started to gain acceptance in 2012. It is expected to create revolutionary changes in the area of e-learning. The system of education, which has necessitated the physical gathering of individuals in the same environment for thousands of years, is about to change. This is because most skills needed in work life may seem too unimportant to be included in the curriculum of institutions providing formal education. Individuals who will demand the training for such as skill may be living in different parts of the world. They may even be in different time zones. The needs that are not seen as values in the classical economics of education reach economic value with the MOOC model with minimal costs.

**Investigating why the MOOC Business Model is successful.**

It is though that the MOOC will be successful. One of the areas with the largest proportional shared in the e-learning market, which is increasing considerably every year, is the MOOC model (Sara Ibn El Ahrache et.al , 2013). Before answering the question of why it is successful, we should first look at significant firms active in this area. The following are such firms.

**Coursera**

Coursera was founded by two academics from Stanford University as a non-profit institution which provides free online education services for everyone's usage. It continuously improves educational materials by making collaborations with universities.

Coursera, which changed its business model in 2013, provides 3 different options for its users. In its business model, it chose to receive income with options of payment per course, course customization and course certificates. However, Coursera also provides free courses, though limited in number.

Considering the profiles of the organizers of the courses uploaded on Coursera, it may be seen that these are usually universities and various organizations. There are 1736 courses on the website by the date 01/12/2016.

Table 1: Coursera (Coursera, 2016)

| Business Model Block | Description |
|---|---|
| Value Proposition | - Courses<br>- Specializations<br>- Online Degrees Academic and technical support<br>- Mobile learning<br>- Sharable Course and Specialization Certificates<br>- An inclusive experience |
| Customer Segment | - Students<br>- Employers<br>- Degree seekers<br>- Knowledge seekers |
| Customer Relationship | - Social Media |
| Channel | - on demand<br>- online |
| Revenue Stream | - Courses<br>- Certificates fee<br>- Specializations |

**Udemy**

It was founded in 2009 by a few entrepreneurs. The content is created by individual educators and marketing partners. Udemy provides various free and paid courses on its website. Certificate opportunities are decided upon by the content creator who prepares the training. As a business model, content creators market their courses to possible customers for a price they determine. Udemy established a business model where it receives a certain amount of commission over the transacted amounts. There are more than 40000 courses by the date 01/12/2016.

Table 2: Udemy (Udemy, 2016)

| Business Model Block | Description |
|---|---|
| Value Proposition | - Courses<br>- Certificates |
| Customer Segment | - Individual users |
| Customer Relationship | - Social Media |
| Channel | - On demand<br>- Affiliates Sales<br>- Ad programs<br>- Coupon Sales |
| Revenue Stream | - Courses |

BEŞTAŞ**Udacity**

It was established in 2012 for profit. The content is created mostly by universities and companies. It has free and paid content. The option of a certificate at the end of the course is only for paid content. There are 155 on the website by 01/12/2016.

Table 3: Udacity (Udacity, 2016)

| **Business Model Block** | Description |
|---|---|
| Value Proposition | Courses |
| | Certificates |
| | Corporate Training |
| | NanoDegree program |
| Customer Segment | Individual Users |
| | Students |
| | Employers |
| Customer Relationship | Social Media |
| Channel | On demand |
| | Ad programs |
| Revenue Stream | Courses |
| | Certificates |
| | NanoDegree program |

**EdX**

It is a non-profit website. The content is created mostly be universities, schools, NGOs and companies. There are free and paid course options on the website. Individuals who want to receive certificates must pay for the certificate.

Table 4: EdX (EdX, 2016)

| Business Model Block | Description |
|---|---|
| Value Proposition | Courses |
| | Certificates |
| | Corporate training |
| Customer Segment | Individual users |
| | Students |
| | Companies |
| | Universities |
| Customer Relationship | |
| Channel | On Demand |
| Revenue Stream | Investments |
| | Sponsorship |
| | Certificate fees |
| | Courses |

The general framework is noticeable in Table 5, where the general classification of business models of firms active in the area of MOOC is given.

Table 5: Classification model for MOOC provider models

| Customer Segment | Students |
| | Employers |
| | Degree seekers |
| | Knowledge seekers |

24

|  | Individual users |
|---|---|
|  | Companies |
|  | Universities |
| Core Value Proposition | Courses |
|  | Specializations |
|  | Online Degrees |
|  | Mobile learning |
|  | Certificates |
|  | Corporate Training |
| Revenue Streams | Courses |
|  | Specializations |
|  | Certificates |
|  | NanoDegree program |
|  | Investments |
|  | Sponsorship |

The Customer Segment category explains the main users of MOOC providers. As seen in Table 5, the most significant customer segment is students, employees and individual users. All these users constitute the main customer layer of MOOC providers. In strategies targeting different customer groups, there are in-service trainings and universities. MOOC providers mainly provide services of courses and certificates. To make the courses more attractive, Coursera provides Specializations service. The received training is supported by Coursera with Online Degree and by Udacity with NanoDegree opportunities.

The main sources of income for MOOC providers are course and certificate fees. They developed Specializations and NanoDegree products to gain an additional income channel. Investments and Sponsorship are important sources of income for the non-profit EdX.

When the MOOC providers were investigated in terms of customer layers and content allocations, the outcome was as it is shown in Figure 1.

Figure 1

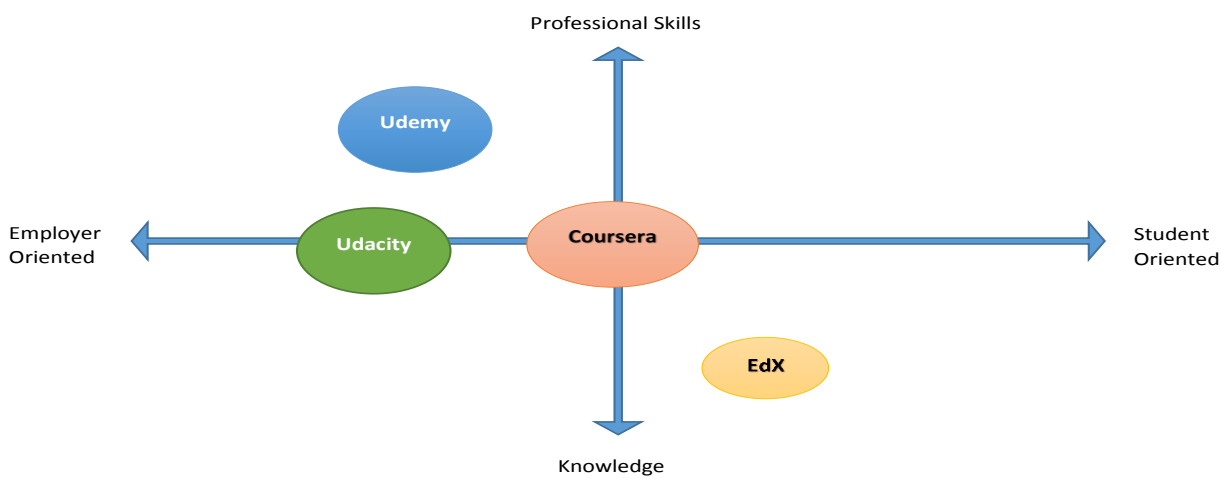

These firms continue to simultaneously provide services for thousands of users demanding education. They are continuously developing their business models to attract individuals. While only videos were shared in the beginning, opportunities like analyses and synthesis were added later.

While collaborations were made with universities for education activities in different fields in the beginning, it was preferred to support the in-service training activities of firms later (CONACHE, DIMA, & MUTU, 2016).

The NanoDegrees model of Udacity started to provide education from the lowest end of the area to an employable point by creating curricula in certain fields. With the specializations business model, Coursera prioritized the education on skills that will be needed more frequently by using curricula in certain fields of expertise. EdX prepared a series of education prepared by experts well-known in their fields with the Xseries model.

While all these started as satisfaction of the needs of individuals who want to receive education on demand, they threaten the classical approach of education by establishing curricula of their own.

Moreover, because some educational needs with normally no economic value and individuals demanding these are scattered around the world as previously stated, this model has an important place in the satisfaction of the needs of individuals who previously could not be attracted by classical educational institutes but constitute large masses all together.

At a point where technology creates change and evolution of an entire human life, it is very problematic that educational methods are still dependent on space. The MOOC model has started to eliminate this physical dependency.

Another important issue in the MOOC model is the educational marketplace business model presented by Udemy. Firms like Udacity, Coursera, EdX moved from the idea that the authority of providing education is for universities, academics or those who made themselves accepted in this area, and provided the preparation of educational material by those defined above. However, Udemy established a marketplace where anyone who has self-confidence can share free or paid content and monetize it. It created a serious difference in the area of MOOCs with this opportunity it provided. This difference is obvious in the number of both its courses and its students.

The MOOC providers Coursera, Udemy, Udacity and EdX provide different opportunities in terms of the cloud computing service type of Platform as a Service. Course lists and course details can be obtained for MOOC providers by API support (Udemy Developer, 2016) (Github EdX, 2016) (Udacity Catalog Api, 2016). EdX provides the highest level of functional support in terms of API support (Github EdX, 2016). The Platform support aimed to be achieved by API support is generally not sufficient.

MOOC providers allow individuals, universities and companies to upload education materials. Then they provide opportunities of forming classrooms and following up on the activities of educational areas. As additional functions, there are quizzes, question and answer sections and certificates. With this aspect, MOOC providers provide SaaS services for those who want to provide education. Their Software as a Service support is strong as there are paid courses for those who want to monetize the educational content they created and there is an opportunity to receive payment.

## CONCLUSION

The firms Udemy, Udacity, EdX and Coursera, which are prominent as MOOC providers, has satisfied the demands of the users with minimal investments with the expandable resources provided by cloud computing. As a result of examining MOOC providers in terms of business models, it was found that their target customers were mostly individual users, students, and employees who require certain skills in professional life. The content target users want to reach is sometimes provided free, while it is sometimes provided for a price. Customization can sometimes be made on paid content based on received demands. To make the provided courses more attractive, there are sometimes supported by online degrees and certificated. Therefore, income is made.

For all business models concerned, it may be seen that change will take place in the future. It was shown that the success of Udemy, which provided content creation and monetization in the form of an educational content marketplace, did not previously reflect the potential of the business models of other sites. Additionally, it was concluded that the provision of more sophisticated skills which will be needed in professional life is the type of content provision needed.

It is seen that the PaaS support of the cloud computing business models of the investigated MOOC providers is still very limited. If this support is developed further, there is a potential of reaching different income channels. In terms of SaaS, while they are still in the initial stages, it was seen that in-service training was provided and fundamental steps were taken for individuals who want to monetize their content.

In general, it is seen that MOOCs are still changing. In the literature review regarding MOOCs, it was seen that 75% of relevant information is about the early stages of MOOCs (Chiappe-Laverde, Hine, & Martínez-Silva, 2015). It is projected that new business models will be tried and MOOCs will change and transform. It is clear that they did not reflect their entire potential so far. It is understood that, as a target audience, MOOCs will be most suitable for individuals who just transformed from students to employees. It is seen that the content will be settled in the form of providing professional skills that are needed dynamically in the workplace.